# Exploring gender differences in the Force Concept Inventory using a random effects meta-analysis of international studies

Purwoko Haryadi Santoso[1,2], Bayu Setiaji[3], Wahyudi[1], Johan Syahbrudin[1], Syamsul Bahri[1,4], Fathurrahman[1,5], A. Suci Rizky Ananda[6], Yusuf Sodhiqin[6]

[1]*Department of Educational Research and Evaluation, Universitas Negeri Yogyakarta, Sleman 55281, Indonesia*
[2]*Department of Physics Education, Universitas Sulawesi Barat, Majene 91413, Indonesia*
[3]*Department of Physics Education, Universitas Negeri Yogyakarta, Sleman 55281, Indonesia*
[4]*Department of Physics Education, Universitas Musamus, Merauke 99611, Indonesia*
[5]*Department of Science Education, Universitas Pendidikan Muhammadiyah Sorong, Sorong 98418, Indonesia*
[6]*Department of Educational Technology, Universitas Negeri Yogyakarta, Sleman 55281, Indonesia*

## Abstract

The force concept inventory (FCI) is one of the research-based assessments (RBAs) established by the physics education research (PER) community to measure students' understanding of Newtonian mechanics. Former works have often recorded the notion of gendered mean FCI scores favoring male students notably in the North America (NA) based studies. Nevertheless, these performance gaps remain inconclusive and unexplored outside the NA context. This paper aims to fill this gap by meta-analyzing the mean FCI scores between gender based on the existing PER literature beyond the NA context. We analyzed the magnitude and direction on the mean FCI scores between gender on the basis of primary international studies published over the last two decades. We also explored the moderating impact of international study characteristics on the meta-analytic findings by performing a subgroup analysis to study the different study regions stratified by two subgroups (NA vs non-NA authors). Thirty-eight studies reporting the mean FCI scores by gender were included in the present meta-analysis. We employed Hedges' *g* statistic to estimate to what degree the mean FCI scores may be different between male and female students on each study. Under a random effects model, we meta-analyzed the findings and conducted a subgroup analysis to answer the research questions. In summary, our meta-analysis indicated a significantly positive and moderate amount of gendered mean FCI scores in favor of male students both in NA- and non-NA based regions, and the performance gaps were wider in the NA-based studies. Suggestions are discussed for promoting gender fairness in the FCI when interpreting its scores for teaching, learning, and forthcoming studies.

## I. Introduction

Assessment is an educational domain that warrants continuous research within the scope of PER [1–4]. One main concern investigated by this subset of PER is designing a valid and reliable research-based assessment (RBA) to gauge students' understanding of physics concepts. Thus far, RBA can be administered as a useful measure to examine some physics learning reforms within the PER. Systematically, the PER community has archived their established RBAs that are openly available from PhysPort [5] or on the LASSO platform [6]. Some of the widely known RBAs published by the community include the Force Concept Inventory (FCI) [7], Force and Motion Conceptual Evaluation (FMCE) [8], Conceptual Survey on Electricity and Magnetism (CSEM) [9], and Brief Electricity and Magnetism (BEMA) [10]. Without omitting the significance of the other RBAs, the focus of this paper is to explore the FCI as one of the most widely implemented RBAs within the community over decades.

The FCI is consistently popular as an RBA that assesses students' conceptual understanding of Newtonian mechanics within PER notably at the undergraduate physics education. Initially, it was designed as 29 multiple-choice conceptual questions [7], which were later revised into 30 items to date [11]. The FCI items were written based on findings from qualitative and quantitative studies about the taxonomy of students' knowledge state in understanding Newtonian mechanics [12]. Since then, many PER scholars have undertaken studies to research its measurement using multiple approaches [13–16]. They also summarized the discovered evidence based on systematic quantitative studies [17,18]. They found that FCI can be acceptable as a useful conceptual metric to measure students' understanding of Newtonian mechanics and to examine the effectiveness of some innovative physics learning strategies developed by PER scholars [19,20]. Recently, exploration has been enriched by utilizing more sophisticated analytical approaches, such as item response theory [21–23], cluster analysis [24], network analysis [25,26], and machine learning [27,28]. Moreover, some studies aimed at validating FCI constructs have been attempted using the framework of factor analysis studies [11,29,30]. Scott and colleagues [31] found that the concept of "Newtonian force" is perfectly valid and that the division of this concept into subcategories by Hestenes and Halloun [7] is also perfectly valid based on their factor analysis findings. Admittedly, FCI is one of the most widely implemented RBAs in producing the milestones of PER studies to date.

According to the lens of social demographic lenses, studying the FCI in terms of gender has attracted much attention in PER works [21,32–34]. Many studies addressing the gender differences[1] in the FCI have been conducted to date [35–41]. Surprisingly, after the FCI was employed as a probe of conceptual understanding, PER scholars discovered that gender discrepancies in mean FCI scores favoring male students may be visible [18,42–44]. One can argue that it is always interesting to gain more knowledge pertaining to this phenomenon as we discover that there are inconclusive findings regarding the magnitude and direction of the gender performance gaps in the FCI from the literature. Some studies suggest that female students were underperformed in the FCI test compared with male students [42,45], while female superiority in FCI was also found elsewhere [46,47]. It is however unclear whether the FCI can significantly affect gendered students' performance in favor of male students [18]. A meta-analysis from Ref [18] has been attempted, but it faces the limitation of only summarizing limited PER literature from the environment in which the original FCI authors are affiliated with the North America (NA) based region[2]. As such, it is critical to update more studies to review the published FCI scores by gender based on the global PER literature beyond the origin of FCI development. This study contributes to filling the void of the former meta-analysis, which is still isolated under the limited environment.

Broadly speaking, the bubble of the FCI development and dissemination has been remarkably endeavored and firstly administered in large scale implementation by North America (NA) PER scholars [21,32,33,42,43,45,48–64]. Nonetheless, Kanim and Cid [65] suggested that the NA-based physics education research generally relies on research subjects who in many respects may not be representative of the global population of physics students. On the other hand, FCI has been widely adopted by PER researchers from other parts of the world internationally across Europe [34,66,67], Asia [46,47,68–76], Australia [77], and Africa [78,79]. These research activities, beyond the NA based regions, can simply denote the recognized reputation of the FCI globally. However, outside the NA context, this has never been explored by the former meta-analysis. Henceforth, a meta-analysis that has been attempted by Ref [18] needs to be updated by reviewing more literature to accommodate the international diversity of the PER literature about the FCI gender performance gaps. The goal intended by this paper is to expand the former meta-analysis study programmed beyond the area of NA-affiliated PER authors.

One can argue that the cultural mechanism that emerged from international publications could give a moderating effect on the gender differences in the mean FCI scores. The performance gaps between gender obtained in the NA-based environment, were often reported in favor of male students [21,32,33,42,43,45,48–64]. Based on the NA literature, female students were also identified as an underrepresented group within the department of physics. By contrast, these performance gaps might be conflicting to the non-NA based studies as reported by Refs [80–82] since the culture of physics students applies differently. In this study, we start to hypothesize that this underexplored phenomena outside the NA context could have a moderating impact on the gendered FCI performance gaps studied in this paper. This hypothesis has been drawn from relevant works of sociological scholars investigating the gender differences seen from between country comparisons [83,84]. In four national contexts, Chan et al. [83] explained that gender differences in physics might be shaped by social identities, social locations, and country contexts from three theoretical ideologies perceived by scientists' attitude toward gender. Their explanation for gender differences in physics was varied by social identities and social location in country specific ways. In mathematics education research, gender differences were also making benefit in favor of male students as reported by Ayalon and Livneh [84]. They revealed that between-country findings in gender performance gaps might be moderated by the different educational system standardization implemented in each country. The use of educational system standardization such as national examinations and less between-teacher instructional variation was evidently a major factor in reducing the advantage of boys over girls in mathematics scores and in the odds of excelling. Drawing on these sociologist findings, instead of merely reporting the common magnitude and direction of the gendered mean FCI scores as combined by a meta-analysis study, this paper will also follow up a subgroup analysis to examine the potential effect of different cultures as represented by distinct study locations to moderate the gendered mean FCI scores summarized by the meta-analytic methodology.

A subgroup meta-analysis is motivated by the prevalence of high heterogeneity between studies included in a meta-analysis. Hence, it should be acknowledged by a choice of the meta-analysis models [85]. Broadly speaking, FCI has been widely recognized within PER community across international study locations. Unique regional characteristics such as demographics, socioeconomic status, cultures, educational systems, and system of beliefs of the samples may inevitably be associated with gendered mean FCI scores summarized by our study. The earlier meta-analysis published by Madsen et al. [18] faces limitation to address the high heterogeneity of the international landscape. They summarized the gendered mean FCI scores based on the weighted average without analyzing the potential heterogeneity of the pooled studies. Ref [18] utilized a fixed effect model$^3$, that is a sort of meta-analysis model when the same true effect size of interest (the FCI gender gap) is supposed to be exactly identical within the pooled studies [86]. In fact, each study must be similar and perform identical methodology then this assumption could be satisfied. Therefore, the fixed-effect model encounters limitations in tackling nonideal conditions when we consider that meta-analysis is summarizing heterogeneous study results [87]. Ignoring heterogeneity leads to an overly precise summary result (that is, the confidence interval is too narrow) and may wrongly imply that a common effect size of interest exists when actually there are real differences in characteristics across studies [88]. To address this limitation, a random effects meta-analysis model is utilized to estimate the summary of gender differences in mean FCI scores based on the international studies.

This study sought to summarize the FCI performance gaps between gender that have been discovered inconclusive and underexplored outside NA based studies. According to the international selection of the existing PER studies beyond the NA context, we start the meta-analysis to summarize the magnitude and direction of the mean FCI scores between gender and subsequently investigate to what degree they can be moderated by a factor of different geographical regions (stratified by two subgroups, NA and non-NA countries). Two research questions are addressed to guide the goal intended by this paper as follows.

RQ1. To what degree are there differences in the magnitude and direction of mean FCI scores between gender?
RQ2. How does the factor of the different study locations (NA vs non-NA studies) moderate the magnitude and direction of mean FCI scores summarized by the present meta-analysis?

The implication of this paper, within its limitations, is exploring to what degree gender differences can be visible by the mean FCI scores reported by a body of published PER literature to date. It should be noted that this paper is unable to establish the conclusion at item level. That FCI gender gaps will indicate the potential of measurement variance must be validated by the lens of psychometric studies such as in differential item functioning (DIF) analysis. Forthcoming study using this novel method must be suggested to enrich the current paper. Replication study is also welcomed to expand the present meta-analytic findings and to summarize the gendered FCI performance gaps based on the more recent and wider published PER literature internationally.

## II. Method

This paper is a meta-analysis study aimed at summarizing the gendered mean FCI scores reported by published PER studies. As a typical systematic review method, the literature search should be initiated in preparing the dataset (a set of literature). This tedious phase was iterated regularly during the study process using the "backward snowballing" technique [85]. To capture the non-NA context, the PER literature should be identified from the international landscape of the FCI administration beyond the NA environment. After the eligible literature has been achieved, we summarized the mean FCI scores between gender rely on the methodology of a meta-analytic review under the random effects model$^3$. Then, we should mitigate the potential of publication bias that might threaten the validity of the meta-analytic conclusion [89]. Subsequently, a subgroup analysis was conducted to examine the moderating effect of different study locations on the gendered mean FCI scores.

## A. Literature search

Broadly speaking, Google Scholar is one of the available indexing databases for systematic review studies [90]. This database platform was chosen since we believed that some studies might be excluded by some subscription-based indexing databases (e.g., Scopus and Web of Science). Then, we gathered our PER literature by scanning through this database using the key queries of "gender" AND "force" AND "concept" AND "inventory". We scrutinized each web page published between 2002 up to 2024 consecutively. The international contexts beyond the circle of NA-affiliated PER scholars were prioritized. In this study, we encompassed the reported mean FCI scores by gender across five authors continents (America, Europe, Asia, Australia, and Africa). Nevertheless, only English-language literature was analyzed in the present meta-analysis study. Meanwhile, we could assure that the readers could replicate the systematic process of our study identification and find an accurate list of the analyzed literature described in the reference section. Furthermore, interested readers could retrace the included literature and replicate what had been analyzed and reviewed throughout our meta-analysis study in this paper.

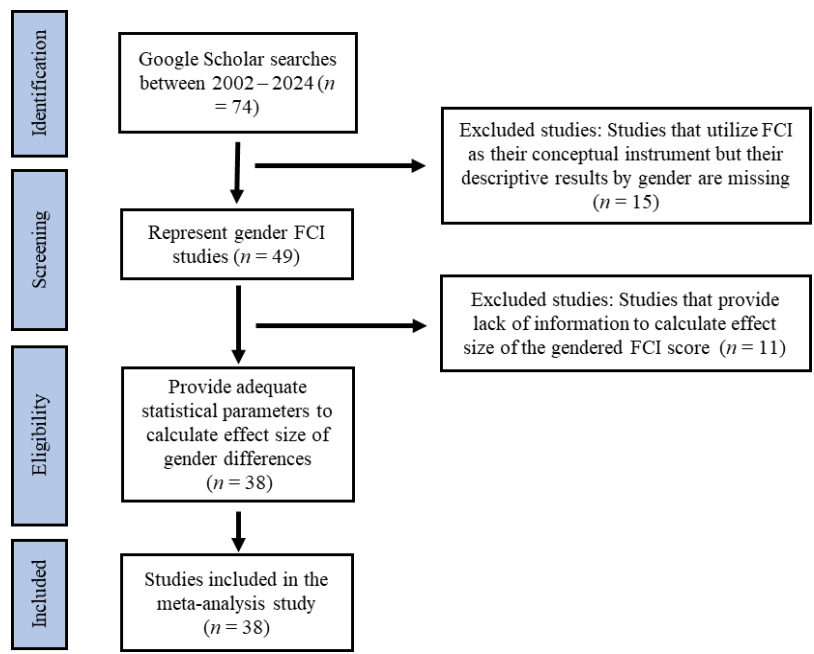

Figure 1. Flow diagram of the inclusion and exclusion criteria for study identification

The inclusion and exclusion criteria for identifying the eligible PER literature in this meta-analysis study can be illustrated in Figure 1. This workflow was prescribed based on the Preferred Reporting Items for Systematic Reviews and Meta-Analyses (PRISMA) guideline [85]. With regard to RQ1, the first inclusion criteria should employ the aforementioned queries of the intended search, and we discovered 74 studies between 2002 and 2024. Subsequently, the exclusion criteria were applied that the eligible studies should explicitly mention the FCI as their measurement tool, and its results should be described by gender (male and female). Admittedly, most of the PER studies often addressed specific subjects, instead of gender studies. Fortunately, they often described the mean FCI scores by gender. In this regard, we then extracted the quantitative findings even though we should manually scrutinize the contents to select studies that clearly provided the gendered comparison of the mean FCI scores. In the screening process, we obtained 49 articles as the prefiltered dataset of our study identification.

The subsequent exclusion setting to filter the eligible studies was the mathematical requirement for the effect size calculation. The prefiltered list of the pooled literature formerly should satisfy the exclusion criteria of the screening process considering the adequacy of statistical parameters needed for the effect size calculation of the FCI gender differences. As described earlier, this study conducted a direct comparison of the mean FCI scores between students' gender binary (male vs female). Therefore, the sort of group comparison effect size method was appropriate based on the purpose of the study intended by our meta-analysis. Three statistical parameters (e.g. sample size ($n$), the mean FCI score ($\bar{x}$), and its standard deviation of gendered group ($s$)) should be explicitly reported by the eligible literature [85]. See equations (1) to (5) below for the mathematical formulas of the "effect size" calculation process. Based on this final exclusion criteria, thirty-eight studies were eligible as our dataset (a set of literature) that would be reviewed and summarized to answer research questions in this paper.

From the eligible PER literature, a pretest-posttest experimental design was found as the most prevalent research design. It should be understood that some studies sometimes provided only pre- or post-instruction data in their findings (e.g. Normandeau et al. [60]). Without omitting the significance of the FCI pretest or posttest data, in this study, we decided to employ the pre-instruction data which usually provided a larger data size in most of the published literature. As such, the gendered scores extracted by the pretest FCI data would be more robust than the post-instruction data with smaller sizes. We assumed that there is no significant difference in the students' preparation between the pretest and posttest session. Hence, the FCI was supposed to be administered within those two different testing conditions equally. Meanwhile, the excluded posttest data were still worth noting to generate our understanding pertaining to the gender differences in the FCI. Admittedly, open room for forthcoming study must be suggested to analyze the post-instruction data. After this selection, we extracted the statistical parameters from the FCI pretest data to estimate the gendered mean FCI scores from the eligible studies.

**B. Quantifying the magnitude of the FCI gender differences from each study: Hedges' $g$ statistic**

The meta-analysis performed in this paper followed the standard procedure of meta-analytic methodology for research synthesis [85]. After the set of eligible literature had been selected ($n = 38$), we formatted a spreadsheet file to tabulate the sample size ($n$), the mean ($\bar{x}$), and the standard deviation ($s$) of the FCI score among male and female groups reported by each literature. To maintain the general scale of the tabulated FCI scores from the pooled literature, the maximum value of the raw FCI score was employed. For studies reporting the FCI scores as a percentage, the linear transformation to the raw score should be calculated before they were used to estimate the gender differences in the mean FCI scores. After that, we quantified the mean differences using Cohen's $d$ [91] to compare the students' performance on the FCI between the male and female groups as follows.

$$d = \frac{\bar{x}_M - \bar{x}_F}{\sqrt{\frac{(n_M - 1)s_M^2 + (n_F - 1)s_F^2}{(n_M - 1) + (n_F - 1)}}} \quad (1)$$

Here, the subscript $M$ indicates the male group and $F$ denotes the female group. Based on equation 1, the direction of the FCI gender differences should be defined by the sign of $d$ value. The positive $d$ value indicates the FCI is obtained higher by the male population.

Then, a measure of discrepancy from the estimated true effect size between studies can be explained by the standard error of Cohen's $d$ ($SE_d$) that is estimated using equation 2 as follows.

$$SE_d = \sqrt{\frac{n_M + n_F}{n_M n_F} + \frac{d^2}{2(n_M + n_F)}} \quad (2)$$

Considering Ref [92], a correction factor is suggested in the case of literature with a small sample size, as indicated by some studies in our pooled literature (e.g. Carleschi et al. [78]). Statistically, Borenstein and colleagues found that it could affect Cohen's $d$ effect size calculation that might be sample dependent [92]. Thus, they proposed to adjust the Cohen's $d$ into Hedges' $g$ value using the correction factor $J$. It can be calculated as follows.

$$J = 1 - \frac{3}{4df - 1} \quad (3)$$

Here, $df$ is the degree of freedom which is equal to $n_M + n_F - 2$. Then, Hedges' $g$ and its standard error of the gender gaps in mean FCI scores are calculated using Equations 4 and 5 as follows.

$$g = J \times d \quad (4)$$

$$SE_g = \sqrt{J} \times SE_d \quad (5)$$

A tabular spreadsheet file was eventually created to organize the list of authors, years, regions, the estimates of the FCI gender differences ($g$) and the standard error ($SE_g$) as our main dataset for meta-analysis. A random effects meta-analysis was estimated using the "`metafor`" package facilitated by the open-source R language [93]. The gender differences in mean FCI scores ($g$) on each study was then labeled based on Cohen's recommendation [91]. A $g$ value that is more than 0.2 should be "weak", greater than 0.5 could be "moderate", and more than 0.8 indicates "strong" differences in the mean FCI scores between male and female students.

As described in section I, this study ought to be novel since we employed the random effects model as our meta-analytic approach. This selection was judged *a priori* by the researchers based on theoretical consequence and the nature of heterogeneity that was evident from the obtained meta-analysis results below [83,84]. Most meta-analyses are based on a collection of studies that may employ unique methods and characteristics of the included literature [93]. It can be assumed that the studies involved in the meta-analysis would not only vary on one of these characteristics but several ones at the same time [89]. Therefore, a random effects model can relax this assumption. Given the expected study diversity, a random effects model would often be more realistic [94]. In this study, a random effects model was estimated using the "`rma.uni`" function from the "`metafor`" package in the R language environment. Mathematical explanation of this model can be found in the Appendix.

Of the thirty-eight studies, they were published by 21 American authors [21,32,33,42,43,45,48–64], 3 European researchers [34,66,67], an Australian scholar [77], 11 Asian researchers [46,47,68–76], and two African scholars [78,79]. Within this global diversity, subsequently, we should consider the need for heterogeneity. The extent to which true effect sizes vary within a set of studies is called between-study heterogeneity. There are three statistical measures to quantify the heterogeneity within the literature. They are the Cochrane's $Q$, $I^2$, and $\tau^2$ statistics [92]. Our null hypothesis of the heterogeneity test statistics was there is no significant difference among the variability of the true effect sizes within the collection of literature. Broadly speaking, when the resulting $p$-value of the heterogeneity test statistics is less than the conventional significance level (e.g. $\alpha = 0.05$), then we should reject the null hypothesis. As such, this could imply that the unexplained heterogeneity is still present from our meta-analytic results. Furthermore, a subgroup analysis is needed to explore the potential moderators such as geographical regions on each study to influence the meta-analyzed gender differences on the FCI scores.

When the high heterogeneity within our pooled literature is present, it could be moderated by the impact of subgroups in our data that have a different true effect size [89,94]. Therefore, we followed up a subgroup analysis in the second research question to account for the unexplained heterogeneity from the meta-analysis findings. In this paper, a subgroup meta-analysis was intended to study the potential effect of different study locations to moderate the identified gender differences in the mean FCI scores. Under the random effects model, a subgroup analysis was conducted to examine the underlying factors moderating the meta-analytic results. Arguably, Ref [89] proposed that a subgroup analysis could be done if the number of studies should be at least 10 in each subgroup. To satisfy this need of covariate distribution [95], we split the dataset as two large subgroups. They were the NA ($n = 21$) and non-NA ($n = 17$) subgroups[2]. This grouping schema should be reasonable in two folds. First, we intended to expand the former meta-analysis in the literature that is heavily focused on analyzing the NA-based studies. Then, there could be interesting results since the FCI gender differences might be present differently in the non-NA based regions which was never explored before. To examine the more diverse cultural contexts beyond the NA based environment, a subgroup analysis should be worth reporting.

**C. The $R^2$ index: an intuitive measure of explained heterogeneity**

The comparison of the subgroups does not inform the amount of heterogeneity that can be explained after we are using the subgroup analysis. To estimate this, Spineli and Pandis [96] propose an index that quantifies the proportion of explained heterogeneity by covariates studied by a subgroup analysis. This index is termed as $R^2$ index, and it is defined as a ratio of the explained heterogeneity to the total heterogeneity as follows.

$$R^2 = 1 - \frac{\tau^2_{unexplained}}{\tau^2_{total}} \quad (6)$$

where $\tau^2_{unexplained}$ is the pooled heterogeneity after the subgroup analysis is conducted and $\tau^2_{total}$ is the heterogeneity that can be accounted before the subgroup analysis is conducted (without splitting the included studies into subgroups under the random effects model). The $R^2$ index will be ranging between 0 and 1 (or in the range 0%-100% if expressed as percentage). The closer the $R^2$ index to 1 or 100% indicates the studied covariates should be determinant affecting the summarized effect sizes in a meta-analysis.

**D. Diagnosing potential of publication bias**

The next our important question warranted by a meta-analysis study under the random-effects model was diagnosing the potential of publication bias. This could imply the imbalance of significant and insignificant results that have been reported by the pooled literature in our dataset. In other words, there might be unpublished results excluded from our dataset (set of literature) [92]. Rosenthal [97] mentioned these unpublished cases as a "file drawer problem", whereas Card [98] termed it as publication bias. Publication bias may threaten the interpretation of the meta-analysis results. A well-written peer-reviewed study could be unpublished in the academic journal due to multiple factors that, for instance the manuscript does not fit with the editorial policy [99]. Although publication bias might be a severe warning to meta-analysis studies, systematic methods for estimating the symmetricity of the eligible published studies could mitigate the risk. In this study, we performed graphical and statistical methods to estimate the publication bias that might be obtained by small size studies. We created a funnel plot, and its symmetricity was statistically tested using Egger's regression and Kendall's rank tests [98].

## III. Results

### A. Descriptive results

We demonstrated that most of the included PER literature in the present meta-analysis was obtained from peer-reviewed papers ($n = 20$) [21,32–34,46,48,54,58,60,61,63,66–71,74,78,79]. Otherwise, they were published in different publication types such as proceeding papers ($n = 9$) [42,43,45,56,57,59,64,73,77], doctoral theses (dissertations) ($n = 7$) [49,50,52,55,62,72,75], and master theses ($n = 2$) [47,76]. Overall, the meta-analysis findings are reported as a forest plot visualized in Figure 2. The Hedges' $g$ (the gender differences in mean FCI scores), the corresponding confidence intervals for each study, and the aggregated estimate of the gendered mean FCI scores with its prediction intervals under the random effects model are presented in Figure 2.

A forest plot is often presented in a typical meta-analysis finding to describe effect sizes on each literature and the combined effect size estimates from the analyzed set of studies. In this paper, complete lists of authors' names, years, and regions are first placed in the leftmost section of Figure 2. The adjacent part from these columns is the estimate of the gender differences in mean FCI scores (effect sizes), and it is visualized by the horizontal bars (95% confidence intervals) with squared dots on each center representing the magnitude of the differences in mean FCI scores between gender within the continuum of observed outcome between -1 and 2. These squared dots are distributed around the dashed line located in the "zero" effect size, indicating that the differences in mean FCI scores between gender should be nonsignificant. The right location of the squared dots from the dashed line describes the outperformed score of the mean FCI scores favoring the male students. By contrast, the left direction of the squared dots from the dashed line suggests that females outperform males in FCI test [46,47]. As we can see, studies with small sample sizes yield greater standard error estimates that could be diagnosed by the wider horizontal bars. Accordingly, the greater size of the squared dots indicate the larger magnitude of the FCI gender gaps on each study. Overall, the common $g$ value is reported on the lowermost of Figure 2 with some heterogeneity statistics ($Q, I^2, \tau^2$). Diamond-shaped dots are given to indicate the summarized value of the gender differences on the mean FCI scores based on our meta-analysis under the random effects model. A prediction intervals is provided accordingly centered at the overall $g$ value to explain the uncertainties of the summary estimate [88].

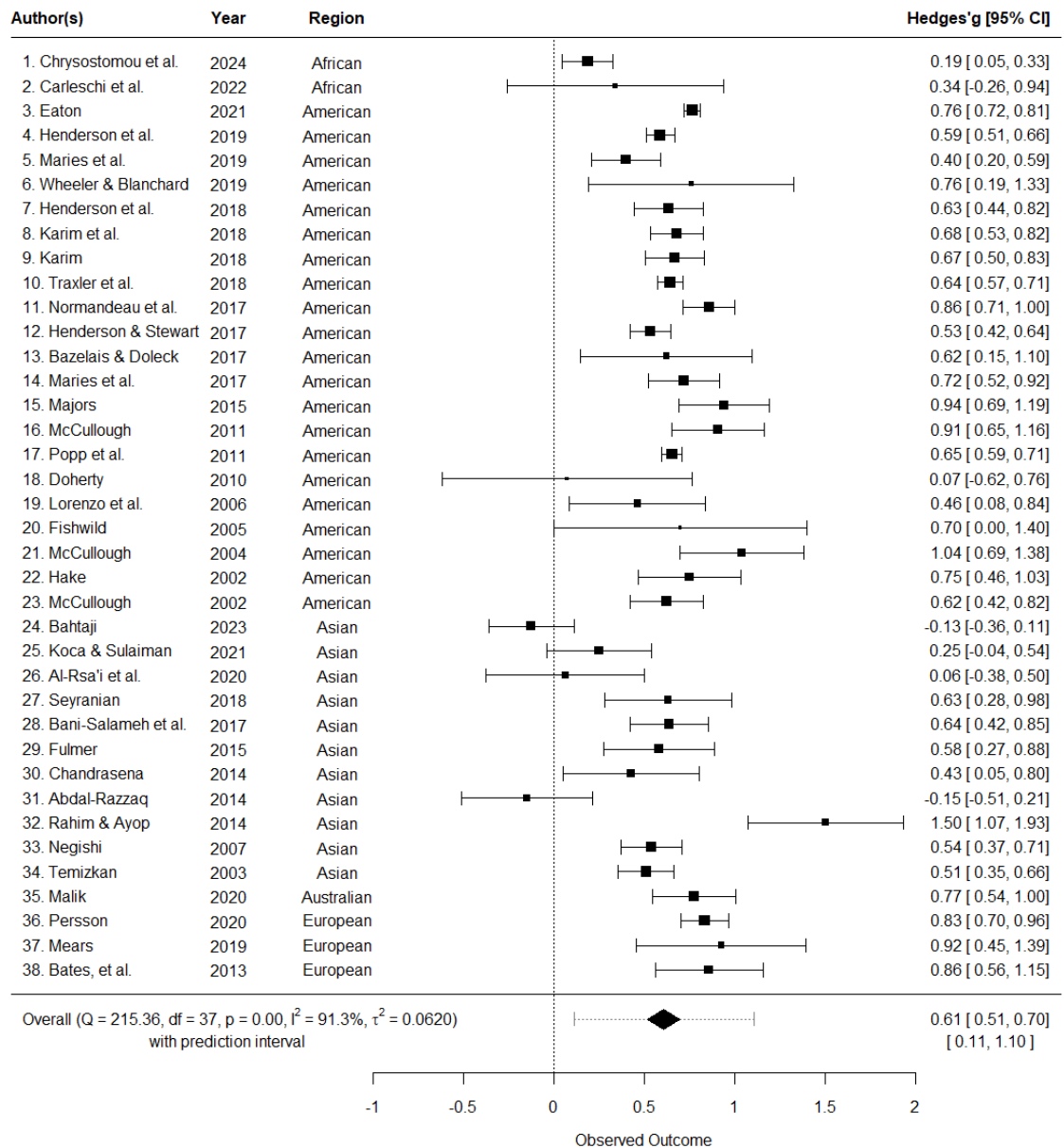

Figure 2. Forest plot showing the results of 38 PER studies examining the gender differences in mean FCI scores. The results are estimated using Hedges' *g* and 95% confidence and prediction intervals.

**B. The magnitude and direction of mean FCI scores between gender (RQ1)**

Thirty-two studies reported significant gender differences on the mean FCI scores in favor of male students. Significant gender differences can be inspected when there is non-zero value within 95% confidence intervals reported by each study (e.g. Chrysostomou [79], Eaton [32], Rahim & Ayop [73]). Overall, the magnitude of the effect sizes (*g*) of the individual studies analyzed in this meta-analysis ranged between -0.15, 95% CI [-0.15, 0.21] [47] and 1.5, 95% CI [1.07, 1.93] [73]. Based on the recommendation of effect sizes labelling suggested by Cohen [91], eight studies should indicate strong effect sizes, most of the pooled studies ($n = 20$) demonstrated moderate effect sizes, and ten studies (e.g., Doherty [49], Al Rsa'I et al. [68], Abdal-Razzaq [47], Bahtaji [46]) indicated weak effect sizes. Most of the literature clearly delineated that the mean FCI score is significantly different between males and females within the PER literature internationally.

Figure 2 is produced under the random effects model of the pooled literature to compute the overall result or summarized effect size (Hedges' $g$). The average gender difference in the mean FCI scores is 0.61, 95% CI [0.51, 0.70] with higher performance in FCI test should be gained by male than female students. This summarized gender difference was significant, $z = 13.0013$, $p < .05$. Then, the prediction intervals of the FCI gender differences generally included non-zero estimate suggesting that future studies could also yield significant results.

Based on the label of Cohen's recommendation, the overall value of FCI gender differences could be a moderate magnitude. We found the positive direction of the gendered mean FCI scores, indicating that male students had obtained higher score in FCI test than female students based on both NA and non-NA literature. This result recommended that male students still tend to achieve a significantly higher FCI score than female students. Most of the included studies ($n = 36$) report the positive direction of Hedges' $g$. Male gender was associated with a greater likelihood of obtaining a higher FCI score. Unsurprisingly, there were two studies from non-NA authors [46,47] showing that their female students outperform the male cohorts as probed by the FCI test.

Indeed, at the end of Figure 2, some statistical measures of heterogeneity are reported. From the $Q$ statistic, it was evident that we should reject the null hypothesis, suggesting that there is a significant heterogeneity of effect size estimates reported by the pooled studies ($p < 0.05$). The test for heterogeneity ($Q = 215.36$, $df = 37$, $p < 0.05$) suggested considerable heterogeneity among the true effects within the pooled studies. In addition, the $I^2$ measure could also quantify whether the heterogeneous factor may be present. The greater $I^2$ value indicated the included literature should be more heterogeneous. In this study, we found $I^2$ value of 91.3%, suggesting a value greater than the cutoff value of 75%, the suggested rule of thumb under which heterogeneous results might be still unexplained across studies. The third measure of residual heterogeneity ($\tau^2$) also demonstrated the similar finding. Our positive $\tau^2$ value was 0.0620, and the rejection of the null hypotheses should be drawn as well. Therefore, our meta-analysis should be followed up further and findings from a subgroup analysis will be reported.

**C. The moderating effect of the different study locations on the gendered mean FCI scores (RQ2)**

We revealed that the heterogeneity should be unavoidable in our pooled literature based on the examination using multiple statistical approaches ($Q, I^2, \tau^2$). Accordingly, this heterogeneity might be accounted by the influence of some possible covariates. Thus, we followed up further analysis to examine the moderating impact driven by different study locations stratified by two subgroups (NA vs non-NA studies). A subgroup analysis was then attempted to further examine the moderating effect. Due to the need of covariate distribution for subgroup analysis as previously described in the method section [89,95], we treated the study locations dichotomously. Both groups were NA ($n = 21$) and non-NA ($n = 17$) based studies. The null hypothesis of this subgroup analysis was that there are no significant differences of the FCI gendered score between NA- and non-NA affiliated studies.

The results of the subgroup analysis are presented in Table 1 and Table 2. Table 1 summarizes the overall heterogeneity test of the subgroup analysis between the NA- and non-NA based studies. In Table 1, we demonstrated that the estimated amount of heterogeneity ($\tau^2$) from the subgroup analysis was equal to 0.0549. Formerly, 0.0620 was the residual heterogeneity ($\tau^2$) obtained from the random-effects model in Figure 2, and 0.0549 was the $\tau^2$ calculated after the subgroup analysis is conducted (Table 1). The $\tau^2$ value in Table 1 (subgroup analysis) was decreasing than $\tau^2$ explained by the random-effects model as described in Figure 2. This reduction indicated the degree of heterogeneity within the pooled literature that could be accounted by the studied covariates (different study locations, NA vs non-NA based literature). The amount of heterogeneity ($R^2$) accounted for can be quantified based on formula explained by Equation 6. It yielded [1 – 0.0549/0.0620] = 11.38%. Using the bootstrapping approach suggested by Viechtbauer [93], one can gauge the precision of the $R^2$ index through the confidence intervals. Based on 95% confidence intervals, we obtained our $R^2$ index was ranging between 0% up to 45.32%. This value was still acceptable in literature [100], and it would quantify to what degree the reduced heterogeneity could be explained by the covariate of the investigated study locations (NA vs non-NA). As many 11.38% of the total amount of heterogeneity could be accounted by including the moderating factor of different study locations into the model.

Table 1. Heterogeneity test of the subgroup analysis among the NA and non-NA study locations

| $\tau^2$ | $p$ | $Q_E$ | $p$ | $Q_M$ | $p$ |
|---|---|---|---|---|---|
| 0.0549 | < 0.05 | 186.1874 | < 0.05 | 189.7808 | < 0.05 |

To assess whether the impact of the study location factor was statistically significant between NA vs non-NA based studies, we reported some examinations of the heterogeneity in Table 1. The notation $\tau^2$ estimates the amount of residual heterogeneity. The notation $Q_E$ describes the test for residual heterogeneity. The notation $Q_M$, as indicated by the subscript, denotes the test for moderators. It was evident that the null hypothesis should be rejected ($Q_M = 189.7808$, $df = 2$, $p < 0.05$), suggesting that investigated different study locations (NA vs non-NA) significantly moderated the gender differences on the mean FCI scores reported by the PER literature globally. Meanwhile, the present subgroup analysis still revealed significant results of the test for residual heterogeneity ($Q_E = 186.1874$, $df = 36$, $p < 0.05$). This ought to be reasonable since one could argue that it was possibly indicating that other moderating factors excluded by our study were influencing the gender differences on the mean FCI scores [93].

Table 2 demonstrates the summary of the gender differences in mean FCI scores separated by two subgroups (NA vs non-NA studies). According to Table 2, both NA and non-NA regions obtained significant estimates of the gender differences on the average FCI scores. The summarized estimate of gender differences recorded by the NA-based literature ($g = 0.6784$, 95% CI [0.56, 0.79]) was larger than that in the non-NA based studies ($g = 0.5092$, 95% CI [0.37, 0.64]). Indeed, there were non-zero values in the confidence intervals of both subgroups. It translates that either NA- based or non-NA based studies produce significant gender gaps on the mean FCI scores. Indeed, we also demonstrated both regions include non-zero value in the estimate of the prediction intervals indicating future studies yielding the same significant gender differences on the gendered mean FCI scores. Clearly, both regions demonstrated the positive direction of effect sizes indicating that the mean FCI scores mostly favors the male students over the female students based on our meta-analysis. Moreover, if this value is labeled based on Cohen's recommendation [91], both subgroups should be categorized as moderate sizes of the gender differences on the mean FCI scores. These labels are showing an immediate similarity with the overall result before the subgroup analysis in Figure 2.

Table 2. Model results of the subgroup analysis by the inclusion of study locations (NA vs non-NA studies)

| Location | Hedges' g | SE | 95% CI | 95% PI | z | p |
|---|---|---|---|---|---|---|
| NA | 0.6784 | 0.0587 | [0.56, 0.79] | [0.21, 1.15] | 11.5597 | < 0.001 |
| Non-NA | 0.5092 | 0.0680 | [0.37, 0.64] | [0.03, 0.98] | 7.4936 | < 0.001 |

Note: CI = confidence intervals, PI = prediction intervals, SE = standard error

**D. Potential of publication bias**

Publication bias could be diagnosed by visualizing the symmetricity of the studies as depicted in a funnel plot (Figure 3). This visually implies how the individual studies scattered around the summary result (the dashed line). Three types of confidence intervals are provided in Figure 3. The significant studies should be located within the 95% confidence intervals indicated by the white, yellow, and pink regions. Thirty-two studies located within this confidence intervals reported significant gender differences in mean FCI scores. It was evident that significant results were mostly reported than the nonsignificant studies in our findings. The *y*-axis is the standard error representing the sample size of the individual studies. The closer value of the standard error to the peak (nearly zero) demonstrates the larger sample size. As we can see, most studies were approaching the apex of the pseudo triangle of confidence intervals in Figure 3. The symmetricity of studies on the left side from the dashed line (summarized estimate) are balanced by those on the right side. Hence, the funnel plot was visually symmetrical. There were 15 study points on the left side from the summary estimate, and the rest of studies were located in the opposite direction. This harmony simply represented no potential publication bias that may be exist in our meta-analysis results.

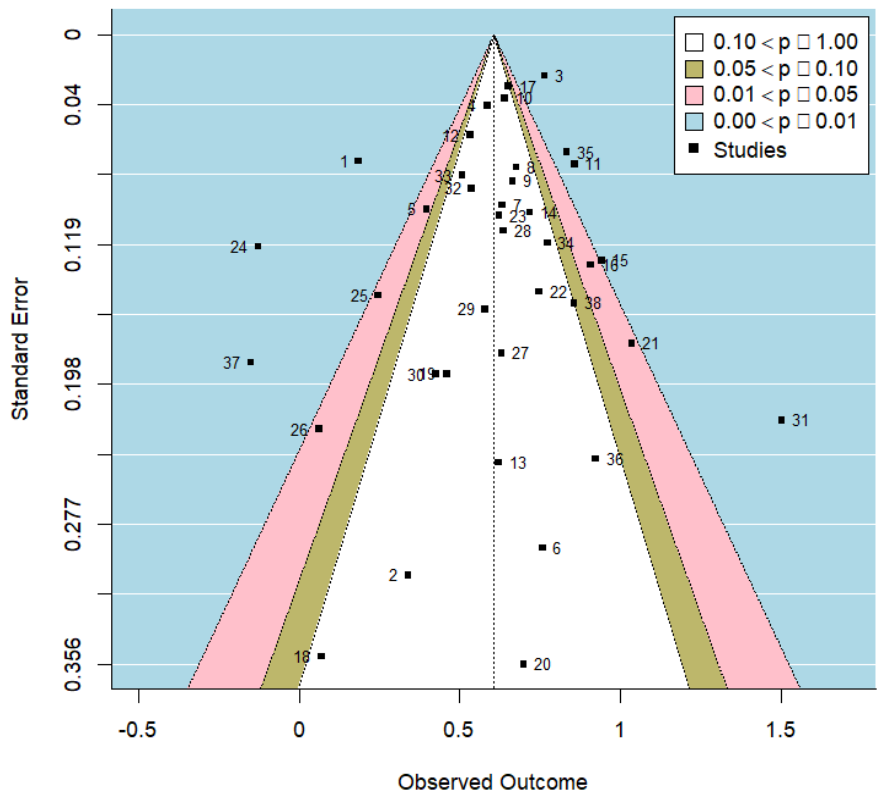

Figure 3. Funnel plot to diagnose the potential of publication bias from the meta-analysis results.

To aid the evidence of the symmetry of the funnel plot (no publication bias evidence), two statistical tests could be employed to examine the symmetricity of the funnel plot. In this study, we utilized Egger's regression and Kendall's tau rank tests [98]. They should add other supports to the previous visual evidence. These statistics test the null hypothesis that the funnel plot is symmetrical. Greater than $\alpha = 0.05$, we obtain nonsignificant results both through Egger's regression test (t = -1.3881, df = 36, p = 0.1736) and Kendall's tau rank test (τ = -0.0242, p = 0.8417). These nonsignificant results suggest that there is no potential bias in our meta-analysis results. Subsequently, we interpret the meta-analytic results to answer the research questions accordingly.

## IV. Discussion

Exploring the FCI in terms of gender has recently attracted many PER scholars to investigate. In this study, we contribute to providing new knowledge within this area. To date, the findings are found inconclusive based on the existing PER literature. A systematic review published by Ref [18] has studied the FCI from the gender perspective, nonetheless, the heterogeneity results among different study locations between NA and non-NA based literature have been underexplored due to fixed effect model is limited to address this concern. In fact, the FCI has gained global recognition within the international PER community beyond the bubble of NA-affiliated environments. One can immediately hypothesize, based on sociological intuition, that the underexplored non-NA context might have moderating impact to the gendered mean FCI scores reported by this paper. Therefore, to address these concerns and to develop the current body of knowledge questioning whether gender discrepancies in favor of male students can be significantly yielded by the FCI measurement, we conduct a meta-analysis study using random effects model. Based on the pooled studies worldwide, our aims are to summarize the magnitude and direction of gender differences that may be present as measured by the mean FCI scores and to study the moderating effect on it influenced by the different study locations.

Based on our meta-analysis of the international PER studies for more than two decades, we obtain that the mean FCI scores are significantly different favoring male students at moderate level ($g = 0.61$, $95\% \ CI$: 0.51, 0.70). Therefore, the male students outperform the female students as reported by most of the studies. This magnitude is significant and the substantial higher FCI scores favoring male students is also discovered from the former meta-analysis work by Ref [18]. Empirically, our meta-analysis demonstrate that this gender performance gaps are still visible as probed by the FCI conceptual assessment based on our literature up to 2024 [46,67,78,79], and we involve global view beyond the NA environment. Nevertheless, it should be noted that the previous meta-analysis [18] reports the different estimate of the gender gaps under the fixed effect meta-analysis model. Thus, the comparison of our findings could be difficult to be done if we solely consider Ref [18].

We demonstrate that high level of heterogeneity is obtained based on the pooled literature included in this paper ($Q(df = 37) = 215.36, \ p < 0.05, \ I^2 = 91.3\%, \ \tau^2 = 0.0620$). As such, we follow up the meta-analysis results in the second research question to consider the geographical stratifications of authors in terms of NA and non-NA based regions. Using a subgroup analysis, we examine the moderating effect of different study locations based on the hypothesis that the different phenomena outside the NA context could have a moderating impact on the gendered FCI performance gaps studied in this paper. Based on our finding, it is evident that the covariate of two geographical groupings (NA vs non-NA) significantly influences the magnitude of the FCI gender gaps. Hence, one can conclude that the magnitude of the gendered mean FCI scores in the non-NA based environment is weaker than in the NA-based regions. Meanwhile, the direction of the gender differences is exactly identical among both regions significantly. Nevertheless, this conclusion is lacking evidence to establish understanding that the FCI is functioning differently between two regions. This should be the next question warranted for forthcoming study. Admittedly, the rationales to explain the different magnitude of the gender performance gaps favoring male students in the non-NA context may be multifactorial. Many sociological mechanisms across countries should be involved to understand this phenomenon and there must be no single factor that is sufficient to explain the gap.

Our hypothesis drawn from sociologist works assumes that different research contexts brought by the non-NA literature should translate the different cultures. This international comparison also may be broadened by assuming physics students across NA vs non-NA groups have different learning atmospheres [83], institutional supports [84], students' backgrounds and preparations, and all possible combinations of many small factors. For example in the different cultures, we can realize that females in the NA-based affiliations are mostly reported as underrepresented groups in the department of physics [18,42–44]. By contrast, the number of female students can be larger within the non-NA environment as reported by Refs [46,47,81,82]. As mentioned above, Abdal-Razzaq [47] reported a competing result with most of the pooled studies. His master's theses using a descriptive quantitative research design discovered that female students could be superior to male engineering students in a classical mechanics course. He concluded that there are no significant gender differences in the FCI. The finding reported by Abdal-Razzaq [47] supports claims suggested by the recent investigation using the qualitative study conducted by Moshfeghyeganeh & Hazari [82] within the context of a Muslim-majority (MM) country. Essentially, both studies are situated within the lower number of male students in the physics class. Moshfeghyeganeh & Hazari [82] articulated the participation of female physicists in the MM countries somewhat different from what was found in the NA-based PER studies. Most of the students involved in the physics class were found to be dominated as described by other study from the Iranian authors [78, 79]. They also captured a typical culture in their country that as many as 60% of women participate in the physics department. Nevertheless, the small amount of heterogeneity accounted by the study locations from our subgroup analysis suggests that the effect of different study locations should be interpreted with caution since the number of nonsignificant studies may be the outliers of our data. In fact, this meta-analysis is only focused on analyzing the average of the FCI score reported by a single study. Therefore, we are limited, or may be out of scope, to examine other mechanisms explained by the social and economic system within a single country to affect the phenomena of gender performance gaps measured by the FCI conceptual assessment. It may be attempted by forthcoming studies to analyze a meta regression analysis to examine some socio-economic variables reported by a single country (for example as reported by Global Gender Gap Report from World Economic Forum[2] that is used to divide the regional subgroups (NA vs non-NA) analyzed in this study).

The strengths of this meta-analysis are the inclusion of the recent published PER studies up to date and the comparison of meta-analysis results between NA- and non-NA studies under the random effects model closing the gap that is still limited from the former work [18]. Many factors can contribute explaining these gender performance differences in physics and many studies has addressed this question to date [101–106]. We admit that there are many important questions that can be sources of uncertainty in our meta-analysis results. Nevertheless, a meta-analysis study is a kind of quantitative research synthesis isolated within the secondary data provided by the published literature. Several drawbacks must be acknowledged. First, high heterogeneity was still present in subgroup analyses that is difficult to avoid in meta-analyses of behavioral studies such as PER [89,94]. Second, some potential factors associated with gender differences as described in the literature such as physics identity [101,102], motivation and belief [103], efficacy [104], personal interest [105], and perceived stereotype [106] are neglected in our meta-analysis. In addition, as previously described in section II, only pre-instruction FCI data are analyzed in the present meta-analysis. Therefore, the findings may be generalized with caution as a global representation of the FCI studies within different testing conditions. We open room for interested readers to broaden the analysis of the current research using the larger FCI data or to synthesize the gender performance differences that can be driven by the other established RBAs within the community.

## V. Conclusion

As a final remark, this meta-analysis has reviewed the gender differences on the mean FCI scores based on the international published PER studies for the recent two decades. A random effects meta-analysis of the eligible PER literature has provided a general finding that a moderate size of the gendered mean FCI score is significantly visible in favor of male students globally. Both NA and non-NA based studies mostly indicate moderate and positive gender performance gaps of FCI scores in average. A covariate of different study locations is evident to significantly moderate the variation of the gendered average FCI scores. Our finding may be unsurprising since the former meta-analysis also achieved the similar conclusion based on the NA literature. Meanwhile, this study has revealed the information from non-NA environment that is still underexplored from the former meta-analysis. Furthermore, our study has implications for researchers and educators to be aware of the prevalence of the performance gaps between students' gender in FCI test. When interpreting the FCI scores, they should consider the magnitude and direction of the discovered performance gaps. Studying the influence of the moderating variables such as diverse contexts as represented by the different students' background (preparation) can deliver useful information to design the more appropriate learning to close the gaps. If the gender differences can be narrowed by revising the FCI constructs among male and female students, PER scholars can attempt further studies to develop the reformed version of the FCI items. A quantitative study intended to equate the reformed version and the original version can be designed to validate the constructs. The equating study is recommended to examine the invariance of the item and person parameters by version of the test that can be estimated using the popular psychometric frameworks within PER such as classical test theory (CTT) and item response theory (IRT). Additionally, implementing innovative physics learning strategies must be enacted to enhance effective student learning regardless of student gender. We admit this meta-analytic review using the random effects model should not be the ultimate conclusion of the notion of the gender differences in FCI. A meta-analysis model is limited to only summarize the average FCI score systematically [85]. One can argue that it may be lacking evidence to summarize the general idea of gender differences from the theory of educational measurement. Admittedly, there are widely known other statistical methods from classical test theory (CTT) and item response theory (IRT) such as differential item functioning (DIF) that can be more robust to establish construct validity of the underlying latent factor of the FCI gender differences [107]. Nevertheless, we need item level data to realize this sort of statistical analysis that is out of scope, but we have to admit that this must be important question for further studies. Oncoming research projects should be proposed to explore other involved sociological or psychological factors that moderate the phenomena and to replicate our gender meta-analysis study to the recent PER literature and the other published RBAs within the PER community.

## Notes

[1] We employ the term "gender differences" to denote the differences of mean FCI scores (or the FCI performance gaps) between male and female students.

[2] In this study, geographical regions are defined based on Global Gender Gap Report 2023 published by World Economic Forum that can be found at https://www.weforum.org/publications/global-gender-gap-report-2023/.

[3] We provide a brief description of fixed effect and random effects meta-analysis models in Appendix. Interested readers are also encouraged to consult the more detailed descriptions in Refs. [85,87,88].

## Acknowledgments

Authors are supported by Beasiswa Pendidikan Indonesia (BPI), a scholarship program for graduate degree which is organized by Balai Pembiayaan Pendidikan Tinggi (BPPT) as a body of Kementerian Pendidikan, Kebudayaan, Riset, dan Teknologi (Kemendikbudristek) and funded by Lembaga Pengelola Dana Pendidikan (LPDP).

## Contributors

PHS and BS conceived the paper. W, JS, SB, and F performed the literature search of the published PER studies between 2002 up to 2024. ASRA and YS collaborated with PHS performing the random-effects meta-analysis model using R language. PHS, BS, W, JS, SB, and F interpreted the results and subsequently discussed them based on their expertise in physics education research. All authors contributed to revising the paper equally.

# Appendices
# Fixed-effect and random-effects meta-analysis model

Meta-analysis is used to synthesize quantitative information from related studies and produce results that summarize a whole body of research [88]. In this study, we used a meta-analytical method to aggregate gender differences in FCI score and obtain a summary estimate based on published PER literature internationally. One important feature of the meta-analysis is its ability to incorporate information about the quality and reliability of the primary studies by weighing larger, better reported studies more heavily [85]. The two quantities of interest are the overall estimate and the measure of the variability in this estimate. Study-level outcomes $\theta_i$ are synthesized as a summarized mean $\hat{\theta}$ according to the study level weights $w_i$:

$$\hat{\theta} = \frac{\sum_i^N \theta_i \cdot w_i}{\sum_i^N w_i} \quad (1)$$

where $i$ is an index of independent study and $N$ is the number of studies. There are two popular statistical models to determine the weight ($w$) for meta-analysis, the fixed effect model and the random effects model [85,87]. These models form the basis for most meta-analyses.

Under the fixed effect model we assume that there is one true effect size $\theta$ that underlies all the studies in the analysis, and thus the observed outcome $Y_i$ for study $i$ is then a function of within-study error $\varepsilon_i$.

$$Y_i = \theta + \varepsilon_i \quad (2)$$

where $\varepsilon_i$ is the difference between the observed mean for study $i$ ($Y_i$) and the common true mean ($\theta$) in which $\varepsilon_i$ is also normally distributed $\varepsilon_i \sim N(0, se(\theta_i)^2)$. In a fixed effect meta-analysis, there is only one source of variation, the estimation error $\varepsilon_i$ or the intrastudy error.

Alternatively, the random effects model supposes that each study samples a different true outcome $\mu_i$, such that the summarized outcome $\mu$ is the grand mean of a population of true effects. Hence, it is possible that all studies share a common effect size, but it is also possible that the effect size varies from study to study. Differences in the methods and sample characteristics may introduce variability (heterogeneity) among the true effects [93]. Random effects models allow us to estimate variation between studies due to differences in methodology, population, sample characteristics, or other factors, thereby allowing a more flexible assessment of similarities or differences between studies [88]. Thus, random effects models are better able to provide a more accurate and representative summary of international FCI scores by gender, despite variations between the studies we used in this analysis. One way to model the heterogeneity is to treat it as purely random. The observed effect $Y_i$ for study $i$ is then influenced by the intrastudy error $\varepsilon_i$ and interstudy error $\xi_i$ [85]. This leads to the random effects model, given by

$$Y_i = \mu + \xi_i + \varepsilon_i \quad (3)$$

where $\xi_i$ is also normally distributed $\xi_i \sim N(0, \tau^2)$, with $\tau^2$ representing the extent of heterogeneity, or between-study (interstudy) variance. $\xi_i$ is the difference between the grand mean ($\mu$) and the true mean for study $i$ ($\theta_i$) and $\varepsilon_i$ is the difference between the true mean of study i ($\theta_i$) and the observed mean for study $i$ ($Y_i$). Therefore, the true effects are assumed to be normally distributed with the mean $\mu$ and variance $\tau^2$. If $\tau^2 = 0$, then it implies homogeneity among the true effects (i.e. $\theta_1 = \theta_2 = \cdots = \theta_N \equiv \theta$), so that $\mu = \theta$ then denotes the true effect. Terms $\mu + \xi_i$ is equal to $\theta$ and equation (3) exactly yields equation (2) accordingly.

Study-level estimates for a fixed effect or random effects model are weighted using the inverse variance:

$$w_i = \begin{cases} \dfrac{1}{se(\theta_i)^2}, & \text{fixed effect} \\ \dfrac{1}{se(\theta_i)^2 + \tau^2}, & \text{random effects} \end{cases} \quad (4)$$

Note that, two formulas described by equation 4 are identical except for the inclusion $\tau^2$ for the random effects model. If the heterogeneity ($\tau^2$) is estimated at zero, then the random effects model collapses to the fixed effect model and the two formulas produce the similar result.